\documentclass[showpacs,twocolumn,nofootinbib]{revtex4}
\usepackage{amsfonts}
\usepackage{graphicx}
\usepackage{amsmath}
\usepackage{amssymb}
\usepackage{amsthm}

\RequirePackage{color}
\RequirePackage[colorlinks=true
,urlcolor=blue
,anchorcolor=blue
,citecolor=blue
,filecolor=blue
,linkcolor=blue
,menucolor=blue
,pagecolor=blue
,linktocpage=true
,pdfproducer=medialab
,pdfa=true
]{hyperref}
\definecolor{blue}{rgb}{0,0,1}
\definecolor{green}{rgb}{0,1,0}
\definecolor{turkos}{rgb}{0,0.4,0.4}
\definecolor{red}{rgb}{1,0,0}
\definecolor{darkblue}{rgb}{0,0,0.5}

\def\be{\begin{equation}}
\def\ee{\end{equation}}
\def\bea{\begin{eqnarray}}
\def\eea{\end{eqnarray}}

\begin{document}

\title{Cosmological constraints on superconducting dark energy models}
\author{Zolt\'an Keresztes$^{1,2}$}
\email{zkeresztes@titan.physx.u-szeged.hu}
\author{L\'{a}szl\'{o} \'{A}. Gergely$^{1,2}$}
\email{gergely@physx.u-szeged.hu}
\author{Tiberiu Harko$^3$}
\email{t.harko@ucl.ac.uk}
\author{Shi-Dong Liang$^{4}$}
\email{stslsd@mail.sysu.edu.cn}
\affiliation{$^{1}$Department of Theoretical Physics, University of Szeged, Tisza Lajos
krt 84-86, Szeged 6720, Hungary, \\
$^2$Department of Experimental Physics, University of Szeged, D\'{o}m T\'{e}%
r 9, Szeged 6720, Hungary,}
\affiliation{$^{3}$Department of Mathematics, University College London, Gower Street,
London, WC1E 6BT, United Kingdom,}
\affiliation{$^4$State Key Laboratory of Optoelectronic Material and Technology, and
Guangdong Province Key Laboratory of Display Material and Technology, School
of Physics and Engineering,\\
Sun Yat-Sen University, Guangzhou 510275, People's Republic of China}

\begin{abstract}
We consider cosmological tests of a scalar-vector-tensor gravitational
model, in which the dark energy is included in the total action through a
gauge invariant, electromagnetic type contribution. The ground state of dark
energy, corresponding to a constant potential $V$ is a Bose-Einstein type
condensate with spontaneously broken U(1) symmetry. In another words dark
energy appears as a massive vector field emerging from a superposition of a
massless vector and a scalar field, the latter corresponding to the
Goldstone boson. Two particular cosmological models, corresponding to pure
electric and pure magnetic type potentials, respectively are confronted with
Type IA Supernovae and Hubble parameter data. In the electric case good fit
is obtained along a narrow inclined stripe in the $\Omega _{m}-\Omega _{V}\,$%
\ parameter plane, which includes the $\Lambda $CDM limit {\color{black} as the best fit}. The other points
on this admissible region represent superconducting dark energy as a sum of
a cosmological constant and a time-evolving contribution. In the magnetic
case the cosmological test selects either i) parameter ranges of the
superconducting dark energy allowing for the standard baryonic plus dark
matter or ii) a unified superconducting dark matter and dark energy model,
additionally including only the baryonic sector.
\end{abstract}

\maketitle


\section{Introduction}

The $\Lambda $CDM ($\Lambda $Cold Dark Matter) cosmological model, in which
our Universe is dominated by two major components, dark energy and dark
matter, has successfully explained most of the recent cosmological
observations. The Planck satellite data from the 2.7 degree Cosmic Microwave
Background full sky survey \cite{P1,P2} provided another important recent
confirmation. The presence of a simple cosmological constant $\Lambda $ in
the Einstein gravitational field equations provides the late-time
acceleration of the Universe \cite{latetime}, a phenomenon representing one
of the most important challenges for present day science. The inclusion of
the cosmological constant in the gravitational field equations however is
highly unsatisfactory, since it lacks a well established physical meaning or
interpretation, and in addition its numerical value cannot be predicted from
present day physical knowledge. From a theoretical point of view it is
desirable to model the mysterious dark energy component of the Universe,
which determines the acceleration of the distant supernovae, either as a
specific physical field, or as a modification of the gravitational force
itself.

An intriguing possibility in this framework is that at large, cosmological
scales the very nature of the gravitational interaction changes, such that
general relativity cannot describe at these scales the dynamical evolution
of the Universe. Several theoretical models, based on various modifications
of general relativity, like $f(R)$ gravity (with $R$ the Ricci scalar) \cite%
{Bu70}, $f\left( R,L_{m}\right) $ type models (with $L_{m}$ the matter
Lagrangian) \cite{Har1}, the $f(R,T)$ modified gravity type models (with $T$
the trace of the energy-momentum tensor) \cite{Har2}, the
Weyl-Cartan-Weitzenb\"{o}ck (WCW) gravity \cite{WCW}, hybrid metric-Palatini
$f(R,\mathcal{R})$ models \textbf{(}$\mathcal{R}$\textbf{\ }being a Ricci
scalar formed from a connection independent of the metric\textbf{) }\cite
{Har3}, $f(R,T,R_{\mu \nu },T_{\mu \nu }$ gravity (with $R_{\mu \nu }$ the
Ricci tensor and $T_{\mu \nu }$ the matter energy-momentum tensor) \cite
{Har4}, or the Eddington-inspired Born-Infeld theory \cite{EIBI} have been
extensively investigated. For a recent review of the generalized
curvature-matter gravitational models of $f\left( R,L_{m}\right) $) and $
f(R,T)$ type see Ref. \cite{Revn}. Modified gravity models can generally
explain the late acceleration of the Universe, without the need of dark
energy, and can also offer some clues on the nature of dark matter.

In a field theoretical approach, among the most important dark energy
candidates are scalar fields, which naturally arise in particle physics and
string theory. Dark energy models involving time-dependent scalar fields
provide a cosmological dynamics in accordance with observations. Moreover,
the early evolution of the Universe in its inflationary era can also be
understood in terms of single or multiple scalar fields, the inflaton
fields, which roll in certain underlying potentials \cite{1}. The action for
gravity and a minimally coupled scalar field in general relativity is \cite%
{Fa04}
\begin{equation}
S_{\phi }=-\int {d^{4}x\sqrt{-g}\left[ \frac{R}{2}-\frac{1}{2}\nabla
^{\alpha }\phi \nabla _{\alpha }\phi +V(\phi )\right] },  \label{acts}
\end{equation}%
with $V(\phi )$ the self-interaction potential of the scalar field. We have
chosen the natural system of units such that the gravitational constant
obeys $G^{-1}=8\pi $, the speed of light $c=1$ and the reduced Planck
constant $\hbar ^{-1}=2\sqrt{\pi }$.

Many different dark energy models were advanced with a scalar field
responsible for the late-time cosmic acceleration. They include those with a
single canonical scalar field $\phi $, the \textit{quintessence} with a
non-zero potential \cite{quintessence}. Another class is provided by the
\textit{$k-$essence} models, in which the late-time acceleration is driven
by the kinetic energy term $X$ of the scalar field and the Lagrangian is an
arbitrary function of $\phi $ and $X$ \cite{kessence}. In its
Dirac-Born-Infeld subcase \cite{tachyon,tachyonSN} $L=-V\left( \phi \right)
\sqrt{1-2X}$, the scalar is known as the tachyonic field. When further the
potential $V$ is a constant, the scalar becomes the Chaplygin gas, also
investigated as a dark energy candidate \cite{Chaplygin}. Some of these
models are also unified models for dark energy and dark matter, similarly to
Refs. \cite{DM_DE}. Finally we mention the coupled models, where dark energy
interacts with dark matter \cite{coupledDE}. For a review of the possible
dark energy candidates see \cite{DEreviews}.

Despite successful scalar field dark energy models, one cannot reject
\textit{a priori} the idea that dark energy has a more complex field
structure. One intriguing possibility is to model dark energy as a vector or
Yang-Mills type field, which may also directly couple to gravity. The
simplest action for a Yang-Mills type dark energy model is \cite{v1}
\begin{eqnarray}
S_{V} &=&-\int d^{4}x\sqrt{-g}\Bigg\{\frac{R}{2}+\sum_{a=1}^{3}\Bigg[\frac{1%
}{16\pi }F_{\mu \nu }^{a}F^{a}{}^{\mu \nu }  \notag \\
&&+V(A^{2})\Bigg]+L_{m}\Bigg\},  \label{actv}
\end{eqnarray}%
where $F_{\mu \nu }^{a}=\nabla _{\mu }A_{\nu }^{a}-\nabla _{\nu }A_{\mu
}^{a} $ and $\nabla _{\mu }$ is the covariant derivative with respect to the
metric, while $A^{2}=g^{\mu \nu }A_{\mu }^{a}A_{\nu }^{a}$. In the action
given by Eq.~(\ref{actv}) dark energy is represented by three vector fields.
This generalizes an Einstein-Maxwell single vector dark energy model, with $%
V(A{}^{2})$ representing a self-interaction potential explicitly violating
gauge invariance. The astrophysical and cosmological implications of the
vector type dark energy models have been extensively investigated in Ref.
\cite{v2}.

One can construct extended vector field dark energy models either, in which
the vector field is non-minimally coupled to the gravitational field. Such
models have been proposed, for example, in Ref. \cite{v3}. The action for
the non-minimally coupled vector dark energy model can be represented as
\begin{eqnarray}
S &=&-\int d^{4}x\sqrt{-g}\Bigg[\frac{R}{2}+\frac{1}{16\pi }F_{\mu \nu
}F^{\mu \nu }-\frac{1}{2}\mu _{\Lambda }^{2}A_{\mu }A^{\mu }  \notag
\label{1} \\
&&+\omega A_{\mu }A^{\mu }R+\eta A^{\mu }A^{\nu }R_{\mu \nu }+L_{m}\Bigg],
\end{eqnarray}%
where $\mu _{\Lambda }$ is the mass of the massive cosmological vector
field, and $A^{\mu }\left( x^{\nu }\right) $, $\mu ,\nu =0,1,2,3$ is its
four-potential, which couples non-minimally to gravity. By analogy with
electrodynamics, the dark energy field tensor is defined as $F_{\mu \nu
}=\nabla _{\mu }A_{\nu }-\nabla _{\nu }A_{\mu }$. Here $\omega $ and $\eta $
are dimensionless coupling parameters.

Inspired by some condensed matter concepts, a so called superconducting type
dark energy model was recently proposed \cite{SupracondDE}. The starting
point of this approach is represented by the deep connection of the
gravitational actions given by Eqs.~(\ref{acts}) and (\ref{actv}),
respectively. Despite their different form, they can be interpreted and
understood as two limiting cases of a single unified fundamental physical
model, describing the spontaneous breaking of the U(1) symmetry of the
"electromagnetic" type dark energy. In condensed matter physics such an
approach is used to describe the superconductive properties of metals from a
field theoretical point of view \cite{W,Cas,F}.

Hence the basic assumption in the superconducting dark energy model is that
its ground state is in the form of a Bose-Einstein type condensate, with the
U(1) symmetry spontaneously broken. Consequently, the U(1) gauge invariant
dark energy appears as a superposition of a vector and of a scalar field,
the latter corresponding to the Goldstone boson. This model is thus
equivalent with the standard field theoretical description of the
superconducting phase. In \cite{SupracondDE} the gravitational field
equations included a non-minimal coupling in the action between the
cosmological mass current and the superconducting dark energy.

Due to the coupling between dark energy and matter current, in the
superconducting dark energy model the total matter energy-momentum tensor $%
^{(m)}T_{\nu }^{\mu }$, including the contributions of both baryonic and
dark matter component, is not conserved. This non-conservation of the matter
energy can be interpreted in terms of the thermodynamics of irreversible
processes as describing matter creation in an open system. However, in the
present model particle creation \textit{is not} the main physical process
determining the accelerated expansion of the Universe.

In the framework of a Friedmann-Robertson-Walker homogeneous and isotropic
geometry two particular cosmological models were considered, corresponding
to the choice of a pure electric type field, and of a pure magnetic type
potential, respectively. In both cases the cosmological evolution was
investigated by numerical and analytical methods. As the main result of
these studies it was shown that in the presence of either the electric or
the magnetic type superconducting dark energy the Universe evolves into an
exponentially accelerating vacuum de Sitter state.

Weather this acceleration emerging in the dark energy models discussed in
Ref. \cite{SupracondDE} falls in the range given by cosmological
observations, still needs to be tested. It is the goal of the present paper
to continue and extend the investigation of the superconducting dark energy
models, by confronting them with the most recent supernovae type Ia (SNIa)
data set \cite{Union21}, then with the available data on the Hubble
parameter-redshift relation \cite{H1,H2,H3}. Such tests are nowadays
routinely applied to all viable dark energy models, hence it is a must to
confront with them the superconducting dark energy models and restrict their
allowed parameter spaces accordingly.

The present paper is organized as follows. In Section~\ref{sup} we briefly
review the mathematical and physical formalism of the superconducting dark
energy model. We perform the cosmological tests in Section~ \ref{Tests}. We
discuss and conclude our results in Section~\ref{concl}.

In what follows we use the Landau-Lifshitz \cite{LaLi} sign conventions for
the definition of the metric and of the geometric quantities.

\section{The superconducting dark energy model}

\label{sup}

In the present Section we will briefly review the gravitational field
equations and the corresponding cosmological application of the
superconducting dark energy model \cite{SupracondDE} to be used in the
sequel. Dark energy is assumed to be described by two fields, a vector field
$A_{\mu }$ (analogous to the electromagnetic 4-potential) and a scalar field
$\phi $ (analogous to the Goldstone boson). Our basic physical requirement
is the breaking of the U(1) symmetry of dark energy by an operator of
"charge" $-2e$, similarly as for superconductivity (see Section 21.6 of Ref.
\cite{W}). Then it follows that the action for a unified scalar-vector dark
energy theory is invariant under the gauge transformations given by $A_{\mu
}(x)\rightarrow A_{\mu }(x)+\partial _{\mu }C(x)$, and $\phi (x)\rightarrow
\phi (x)+C(x)$, where $C(x)$ is an arbitrary function of the coordinates.

Hence U(1) gauge invariance of the dark energy requires that the fields $%
\left( \phi ,A_{\mu }\right) $ should appear only in the combinations $%
F_{\mu \nu }=\nabla _{\mu }A_{\nu }-\nabla _{\nu }A_{\mu }$ and 
\begin{equation}
\mathfrak{A}_{\mu }=A_{\mu }-\nabla _{\mu }\phi ~,
\end{equation}
 respectively.

\subsection{Action and gravitational field equations}

An action for the superconducting dark energy model was proposed in Ref.
\cite{SupracondDE}. We modify that action here such that it becomes
manifestly gauge-invariant, hence the vector and the scalar fields appear
only in the combinations $F_{\mu \nu }$ and $\mathfrak{A}_{\mu }$. The
action is:
\begin{eqnarray}
S &=&-\int d^{4}x\sqrt{-g}\Bigg[\frac{R}{2}+\frac{1}{16\pi }F_{\mu \nu
}F^{\mu \nu }-\frac{\lambda }{2}\mathfrak{A}^{2}+V\left( \mathfrak{A}%
^{2}\right)  \notag \\
&&-\frac{\alpha }{2}g^{\mu \nu }j_{\mu }\mathfrak{A}_{\nu }+L_{m}\left(
g_{\mu \nu },\psi \right) \Bigg],  \label{s1}
\end{eqnarray}%
where $\mathfrak{A}^{2}=g^{\mu \nu }\mathfrak{A}_{\mu }\mathfrak{A}_{\nu }$,
the notation $V\left( \mathfrak{A}^{2}\right) $ stands for the
self-interaction potential of the $\mathfrak{A}_{\mu }$ field, $\lambda =$ $%
\mu _{\Lambda }^{2}/2$ is related to the mass of the massive cosmological
vector field, while $L_{m}\left( g_{\mu \nu },\psi \right) $ is the
Lagrangian of the total (baryonic plus dark) matter, represented by $\psi $.
In Eq.~(\ref{s1}) $j^{\mu }=\rho u^{\mu }$ denotes the total cosmological
mass current, where $\rho $ is the total (baryonic plus dark) matter
density, and $u^{\mu }$ is the matter four-velocity. We assume that the
baryonic and dark matter are comoving with the cosmological expansion. In
the action (\ref{s1}) we have also introduced an interaction term between
the total matter flux $j^{\mu }$ and the superconducting dark energy gauge
invariant potentials $A_{\mu }-\nabla _{\mu }\phi $, with $\alpha $
representing a coupling constant.

The action (\ref{s1}) has two important limiting cases. When $\phi \equiv 0$%
, and $\lambda =0$, $\alpha =0$, the gravitational action (\ref{s1}) gives
the single field vector model of dark energy \cite{v1,v2}, a subcase of the
action (\ref{actv}). On the other hand, when the electromagnetic type
potential vanishes, $A^{\mu }=0$, and $\lambda =1$, $V\left( \mathfrak{A}%
^{2}\right) =0$, we reobtain the gravitational action of the minimally
coupled scalar-tensor theory~(\ref{acts}), for $V(\phi )=0$. Therefore the
gravitational action (\ref{s1}) provides a unified framework for the
inclusion into the action of a dark energy theory of the minimal
scalar-vector interactions, under the fundamental assumption of the
existence of a U(1) type broken symmetry. It is interesting to note that the
second and third terms in Eq.~(\ref{s1}) are similar to the corresponding
terms in the Stueckelberg Lagrangian \cite{Stue}. Indeed, massive vector
fields replace a massless vector field and a Goldstone boson in the
nonrelativistic treatment of superconductivity by Anderson \cite{Anderson}
and the related Englert-Brout-Higgs-Guralnik-Hagen-Kibble mechanism of mass
generation by spontaneous symmetry breaking \cite{EBHGHK}.

By varying the action (\ref{s1}) with respect to the metric tensor, the
scalar field $\phi $ and the vector potential $A_{\mu }$ we obtain the full
set of the gravitational field equations for the superconducting dark energy
model as
\begin{eqnarray}\label{fn1}
&&\hspace*{-5mm}R_{\mu \nu }-\frac{1}{2}Rg_{\mu \nu }=T_{\mu \nu }+\frac{1}{%
4\pi }\left( -F_{\mu \alpha }F_{\nu }^{\alpha }+\frac{1}{4}F_{\alpha \beta
}F^{\alpha \beta }g_{\mu \nu }\right) +  \notag  \label{feq} \\
&&\hspace*{-5mm}\lambda \mathfrak{A}_{\mu }\mathfrak{A}_{\nu }-\frac{\lambda
}{2}\;\mathfrak{A}^{2}g_{\mu \nu }+\alpha j_{\mu }\mathfrak{A}_{\nu }-\frac{%
\alpha }{2}j^{\beta }\mathfrak{A}_{\beta }g_{\mu \nu }-2\times  \notag \\
&&\hspace*{-5mm}\left[ \partial _{\mathfrak{A}^{2}}V\left( \mathfrak{A}%
^{2}\right) \mathfrak{A}_{\mu }\mathfrak{A}_{\nu }-\frac{1}{2}V\left(
\mathfrak{A}^{2}\right) g_{\mu \nu }\right] ,
\end{eqnarray}%
\begin{eqnarray}\label{fn2}
&&\lambda g^{\mu \nu }\nabla _{\mu }\nabla _{\nu }\phi +2g^{\mu \nu }\nabla
_{\nu }\left[ \partial _{\mathfrak{A}^{2}}V\left( \mathfrak{A}^{2}\right)
A_{\mu }\right] -  \notag \\
&&\lambda \nabla _{\mu }A^{\mu }-\frac{\alpha }{2}\nabla _{\mu }j^{\mu }=0,
\end{eqnarray}%
and
\begin{equation}
\frac{1}{4\pi }\nabla _{\nu }F^{\mu \nu }=J^{\mu },  \label{DivF}
\end{equation}%
respectively, where we have denoted
\begin{equation}
J^{\mu }=\left[ \lambda \mathfrak{A}^{\mu }+\frac{\alpha }{2}j^{\mu
}-2\partial _{\mathfrak{A}^{2}}V\left( \mathfrak{A}^{2}\right) A^{\mu }%
\right] .  \label{j}
\end{equation}%
In order to obtain Eqs.~(\ref{feq})-(\ref{j}) we have used the mathematical
identities
\begin{eqnarray*}
&&\frac{\delta V\left( g^{\mu \nu }\mathfrak{A}_{\mu }\mathfrak{A}_{\nu
}\right) }{\delta g^{\mu \nu }}=\partial _{\mathfrak{A}^{2}}V\left(
\mathfrak{A}^{2}\right) \mathfrak{A}_{\mu }\mathfrak{A}_{\nu }\delta g^{\mu
\nu }, \\
&&\frac{\delta V\left( g^{\mu \nu }\mathfrak{A}_{\mu }\mathfrak{A}_{\nu
}\right) }{\delta \phi }=-2\partial _{\mathfrak{A}^{2}}V\left( \mathfrak{A}%
^{2}\right) g^{\mu \nu }A_{\mu }\nabla _{\nu }\delta \phi , \\
&&\frac{\delta V\left( g^{\mu \nu }\mathfrak{A}_{\mu }\mathfrak{A}_{\nu
}\right) }{\delta A_{\mu }}=2\partial _{\mathfrak{A}^{2}}V\left( \mathfrak{A}%
^{2}\right) g^{\mu \nu }A_{\nu }\delta A_{\mu },
\end{eqnarray*}%
and
\begin{eqnarray*}
&&-2\int \left[ \partial _{\mathfrak{A}^{2}}V\left( \mathfrak{A}^{2}\right)
g^{\mu \nu }A_{\mu }\nabla _{\nu }\delta \phi \right] \sqrt{-g}d^{4}x= \\
&&2\int g^{\mu \nu }\nabla _{\nu }\left[ \partial _{\mathfrak{A}^{2}}V\left(
\mathfrak{A}^{2}\right) A_{\mu }\right] \sqrt{-g}d^{4}x,
\end{eqnarray*}%
respectively, and we have assumed that at infinity the variation of the
scalar field $\phi $ vanishes.

In the following we assume that the matter content of the Universe consists
of a perfect fluid, and therefore the energy-momentum tensor $^{(m)}T_{\mu
\nu}$ of the matter, including both the dark matter and the baryonic matter
components, with energy densities and pressures $\rho _{DM}$, $p_{DM}$ and $%
\rho _B$ and $p_B$ respectively, is given by
\begin{equation}  \label{9}
^{(m)}T_{\mu \nu}=\left(\rho +p\right)u_{\mu}u_{\nu}-pg_{\mu \nu},
\end{equation}
where $\rho =\rho _{DM}+\rho _B$ and $p=p_{DM}+p_B$ are the total
thermodynamic energy density and pressure of the matter components (baryonic
and dark). The dark energy electromagnetic type tensor $F^{\mu \nu }$
automatically satisfies the Bianchi identity
\begin{equation}
\varepsilon ^{\alpha \beta \mu \nu }\nabla _{\beta }F_{\mu \nu }=0,
\end{equation}
where $\varepsilon ^{\alpha \beta \mu \nu }$ is the complete antisymmetric
unit tensor of rank four.

The conservation equation, including all the components of the Universe in
the presence of the superconducting dark energy is obtained as
\begin{eqnarray}
&&\hspace*{-9mm}\nabla _{\mu }\;^{(m)}T_{\nu }^{\mu }+\frac{\alpha }{2}%
\nabla _{\mu }\left[ j^{\mu }\mathfrak{A}_{\mu }\right] -\frac{\alpha }{2}%
\nabla _{\nu }j^{\beta }\mathfrak{A}_{\beta }+  \notag \\
&&\hspace*{-9mm}2\nabla _{\mu }\nabla _{\nu }\left[ \partial _{\mathfrak{A}%
^{2}}V\left( \mathfrak{A}^{2}\right) A^{\mu }\right] +2\partial _{\mathfrak{A%
}^{2}}V\left( \mathfrak{A}^{2}\right) A^{\alpha }\nabla _{\alpha }A_{\nu }=0.
\label{conservation}
\end{eqnarray}

In the rest of the paper we restrict our analysis to the case of a constant
self-interaction potential of the superconducting dark energy field, $%
V\left( \mathfrak{A}^{2}\right) =V_{0}=\mathrm{constant}$. It is interesting
to note that in the theory of superconductors such a potential corresponds
to the average energy of the ground state of the condensate $<\epsilon
_{\alpha \beta }\psi ^{\alpha }\psi ^{\beta }>$ formed in the
superconductor, with $\psi ^{\alpha }$ the wave function of the condensate,
and $\epsilon _{\alpha \beta }$ the two-dimensional completely antisymmetric
unit tensor \cite{W,Cas}.

\subsection{Cosmological models}

For cosmological applications we assume that the metric of the Universe is
given by the isotropic and homogeneous flat Friedmann-Robertson-Walker
metric,
\begin{equation}
ds^{2}=dt^{2}-a^{2}(t)\left( dx^{2}+dy^{2}+dz^{2}\right) ,
\end{equation}%
where $a(t)$ is the scale factor describing the expansion of the Universe.

In the following we will discuss two types of superconducting dark energy
models, the electric and magnetic type models, respectively.

\subsubsection{The electric type (scalar) superconducting dark energy model}

In the electric type (scalar) dark energy model we assume that the dark
energy vector potential has the simple form $A_{\mu }=\left(
A_{0}(t),0,0,0\right) $, hence only the electric (scalar) potential is kept.
This dark energy 4-vector potential leads to $F_{\mu \nu }\equiv 0$, $%
\forall \mu ,\nu \in \lbrack 0,1,2,3]$, hence both the "electric" and the
"magnetic" dark energy fields vanish and the scalar $A_{0}(t)$ turns out to
be non-dynamical in a Lagrangian sense. In other applications of the
superconducting dark energy models however, notably in the static,
spherically symmetric case, when $A_{0}=\phi (r)$, a pure electric type
field is generated. That case is important when discussing the Newtonian
limit, or the dark energy effects on the planetary dynamics and light motion
in the Solar System. Therefore we adopt the unique terminology "electric
type superconducting dark energy model" for this case.

Since $F_{\mu \nu }=0$, we have $J^{\mu }=0$ from Eq. (\ref{DivF}), then Eq.
(\ref{j}) gives
\begin{equation}
A_{0}-\dot{\phi}=-\frac{\alpha \rho }{2\lambda }  \label{coupling}
\end{equation}
for a constant potential $V\left( \mathfrak{A}^{2}\right) =V_{0}$. Only the
combination $A_{0}-\dot{\phi}$ occurs in the field equations which can be
eliminated by $-\alpha \rho /\left( 2\lambda \right) $ \cite{SupracondDE}.
Hence the generalized Friedmann equations describing the cosmological
expansion in the electric type superconducting dark energy model become
\begin{equation}
3H^{2}=\rho -\frac{\alpha ^{2}}{8\lambda }\rho ^{2}+V_{0}=\rho +\rho _{DE}~,
\label{hd0}
\end{equation}%
\begin{equation}
2\dot{H}+3H^{2}=-p+\frac{\alpha ^{2}}{8\lambda }\rho ^{2}+V_{0}=-p-p_{DE}~,
\label{pd0}
\end{equation}%
where we have denoted
\begin{equation}
\rho _{DE}=-\frac{\alpha ^{2}}{8\lambda }\rho ^{2}+V_{0},p_{DE}=-\frac{%
\alpha ^{2}}{8\lambda }\rho ^{2}-V_{0}~,
\end{equation}
{\color{black} and $H=\dot{a}/a$ is the Hubble parameter. From Eqs. (\ref{hd0}
), then Eq. (\ref{pd0}), we have
\begin{equation}
2\dot{H}=-\rho \left( 1+\frac{p}{\rho }-\frac{\alpha ^{2}}{4\lambda }\rho
\right) ~.  \label{EHdot}
\end{equation}
Then from the derivative of Eq. (\ref{hd0}) with respect to time and
eliminating $\dot{H}$ by Eq. (\ref{EHdot}), we find
\begin{equation}
\dot{\rho}\left( 1-\frac{\alpha ^{2}}{4\lambda }\rho \right) +3H\rho \left(
1+\frac{p}{\rho }-\frac{\alpha ^{2}}{4\lambda }\rho \right) =0~.
\label{Erhobalance}
\end{equation}

For dust ($p=0$) this reduces to
\begin{equation}
\left( \dot{\rho}+3H\rho \right) \left( 1-\frac{\alpha ^{2}}{4\lambda }\rho
\right) =0~,
\end{equation}
giving a non-constant solution} $\rho \propto a^{-3}$ expressed in terms of
redshift ($z=a_{0}/a-1$):
\begin{equation}
\rho (z)=\rho _{0}\left( 1+z\right) ^{3},  \label{fit1}
\end{equation}
where the subscript $0$ denotes the present value. Finally the Hubble
parameter is given by
\begin{equation}
3H^{2}(z)=\rho _{0}(1+z)^{3}-\frac{\alpha ^{2}\rho _{0}^{2}}{8\lambda }%
(1+z)^{6}+V_{0}~.  \label{fit2}
\end{equation}
{\color{black} Despite of $\alpha \neq 0$, the four-divergence of the dust's
energy-momentum tensor vanishes in Eq. (\ref{conservation}), and the matter
density satisfies a continuity equation. However {a non-vanishing coupling
between the matter and the fields }$A$ and $\phi $ leads to the excitations
of the dark energy as can be seen from Eq. (\ref{coupling}).}

\subsubsection{Magnetic (vector type) superconducting dark energy models}

In the following we call magnetic type superconducting dark energy models
the cosmological models in which the 4-vector potential is given by $A_{\mu
}=\left( 0,A_{1}(t),A_{2}(t),A_{3}(t)\right) $, hence by the magnetic
(3-vector) potential. For an isotropic expansion the components of $A_{\mu }$
satisfy the condition $A_{1}(t)=A_{2}(t)=A_{3}(t)=A(t)$. However, this model
has the interesting property that because of the choice of the potential and
of the space-time metric, and its symmetry properties, \textit{the "magnetic
field" identically vanishes}, but \textit{the electric field is non-zero}.
Again we note that in the static, spherically symmetric case a proper dark
energy magnetic field emerges from the vector potential $\vec{A}=\vec{A}%
\left( \vec{r}\right) $. Therefore, for the uniqueness of the terminology,
we named this case according to the magnetic type potential as magnetic type
(vector) superconducting dark energy model.

With the constant value $V\left( \mathfrak{A}^{2}\right) =V_{0}$ {\color{black}
and homogeneous, isotropic 4-vector potential, the field equation (\ref%
{DivF}) gives $J^{t}=0$ which due to Eq. (\ref{j}) leads to
\begin{equation}
\dot{\phi}=\frac{\alpha \rho }{2\lambda }~.  \label{coupling2}
\end{equation}%
By this expression $\dot{\phi}$ can be eliminated from the remaining
equations. The} generalized Friedmann and the energy conservation equations
take the form
\begin{equation}
3H^{2}=\rho -\frac{\alpha ^{2}}{8\lambda }\rho ^{2}+\frac{3}{8\pi }\frac{%
\dot{A}^{2}}{a^{2}}+\frac{3\lambda }{2}\frac{A^{2}}{a^{2}}+V_{0}~,
\label{52}
\end{equation}%
\begin{equation}
2\dot{H}+3H^{2}=-p+\frac{\alpha ^{2}}{8\lambda }\rho ^{2}-\frac{1}{8\pi }%
\frac{\dot{A}^{2}}{a^{2}}+\frac{\lambda }{2}\frac{A^{2}}{a^{2}}+V_{0}~,
\label{53}
\end{equation}%
\begin{equation}
\ddot{A}+H\dot{A}+4\pi \lambda A=0~.  \label{54n}
\end{equation}%
{\color{black} From Eqs. (\ref{52}) and (\ref{53}) we have
\begin{equation}
2\dot{H}=-\rho \left( 1+\frac{p}{\rho }-\frac{\alpha ^{2}}{4\lambda }\rho
\right) -\frac{1}{2\pi }\frac{\dot{A}^{2}}{a^{2}}-\lambda \frac{A^{2}}{a^{2}}%
~.  \label{BHdot}
\end{equation}%
Then from the time-derivative of (\ref{52}) and eliminating $\dot{H}$ and $%
\ddot{A}$ by Eqs. (\ref{BHdot}) and (\ref{54n}), respectively, we find the
same evolution equation (\ref{Erhobalance}) for the matter density than
in the electric-type dark energy model case.

The dust satisfies a matter conservation separately, however Eq. (\ref%
{coupling2}) tells that the matter is a source of the scalar field $\phi $ for a
non-vanishing value of $\alpha $ .}

\section{Testing the superconducting dark energy model with supernovae and
Hubble parameter data sets \label{Tests}}

The confrontation of a cosmological model with the Union 2.1 SNIa data set
\cite{Union21} through a $\chi ^{2}$-test, as described in Refs. \cite%
{tachyonSN} relies on fitting with the computed luminosity distance ($d_{L}$
)-redshift relation. For a flat Friedmann universe a dimensionless
luminosity distance $\hat{d}_{L}=H_{0}d_{L}$ is given by
\begin{equation}
\frac{d}{dz}\frac{\hat{d}_{L}}{1+z}=\frac{1}{\hat{H}}~,  \label{dLz}
\end{equation}%
where $z$ is the redshift, $H\left( z\right) $ is the Hubble parameter, $%
H_{0}=H\left( z=0\right) $ and $\hat{H}\left( z\right) =H\left( z\right)
/H_{0}$. Hereafter the subscript $0$ always denotes the present value of the
corresponding quantity.

Using the data set \cite{H1,H2,H3} on the Hubble parameter-redshift relation
we compute%
\begin{equation}
\chi _{H}^{2}=\sum_{i}\frac{\left[ H_{th}\left( z_{i}\right) \!-\!\overline{H%
}_{obs}\left( z_{i}\right) \right] ^{2}}{\sigma _{i}^{2}}~,  \label{chiH}
\end{equation}%
where $H_{th}\left( z_{i}\right) $\ and $\overline{H}_{obs}\left(
z_{i}\right) $\ are the Hubble parameter values at redshifts $z_{i}$,
predicted by the cosmological model, and determined from the observations,
respectively, while $\sigma _{i}$ is the scattering in $\overline{H}%
_{obs}\left( z_{i}\right) $. We note that a subset of the data set on the
Hubble parameter-redshift relation was used to emphasize a tension with the $%
\Lambda $CDM model (by computing the two-point $Om h^{2}$ function) \cite%
{SSS}. In the following we perform a test of the superconducting dark energy
model with the combined SNIa and Hubble parameter data set by calculating $%
\chi _{SNIa+H}^{2}=\chi _{SNIa}^{2}+\chi _{H}^{2}$, where $\chi _{SNIa}^{2}$%
\ is the $\chi ^{2}$-value from the confrontation with the SNIa data set.

In our analysis we will investigate only the late cosmological evolution.
Then the mixture of the baryonic and the dark matter component has
negligible pressure, so that in the field equations we can take $p=0$.

\subsection{Cosmological tests of the electric dark energy model}

\begin{figure}[tbp]
\includegraphics[width=8.5cm, angle=0]{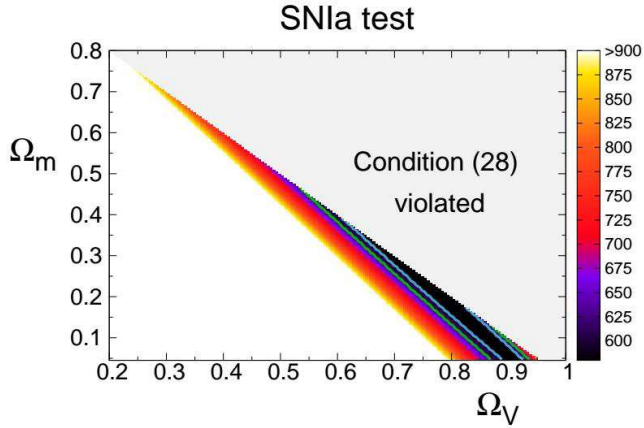}
\caption{(Color online) The test of the electric dark energy model with the
supernovae data set is shown. The color scale represents the $\protect\chi %
^{2}$ values on the parameter space ($\Omega _{V}$,$\Omega _{m}$). Within
the gray area the condition (\protect\ref{cond1}) is violated. The
light-blue and the green lines bounds the 1$\protect\sigma $ and 2$\protect%
\sigma $ confidence regions, respectively.}
\label{TestESDESNIa}
\end{figure}

In this model the dark energy vector potential $A_{\mu }$ is parallel with
the four velocity $u_{\mu }$ of the comoving matter. The evolution of the
Hubble parameter follows from Eqs.~(\ref{fit1}) and (\ref{fit2}), and is
given by
\begin{equation}
\hat{H}^{2}\left( z\right) =\Omega _{m}\left( 1+z\right) ^{3}+\left( \Omega
_{DE}-\Omega _{V}\right) \left( 1+z\right) ^{6}+\Omega _{V}~,  \label{HzESDE}
\end{equation}%
where $\Omega _{m}$ and $\Omega _{DE}$ are the present day density
parameters of the dust and dark energy, respectively:%
\begin{equation}
\Omega _{m}=\left. \frac{\rho }{3H^{2}}\right\vert _{z=0}~,
\label{mattparam}
\end{equation}%
\begin{equation}
\Omega _{DE}=1-\Omega _{m}=-\widetilde{\alpha }\Omega _{m}^{2}+\Omega _{V}~,
\end{equation}%
and%
\begin{equation}
\widetilde{\alpha }=\frac{3\alpha ^{2}H_{0}^{2}}{8\lambda }~,~\Omega _{V}=%
\frac{V_{0}}{3H_{0}^{2}}~,
\end{equation}%
are dimensionless constants. We constrain the model parameters by imposing
that the universe continuously expanded after the Big Bang Nucleosynthesis
(BBN). Since $\hat{H}^{2}\left( z\right) \geq 0$, we have the following
constraint on the parameter space,
\begin{equation}
\widetilde{\alpha }<\widetilde{\alpha }_{cr}=\frac{\left( 1+z_{BBN}\right)
^{3}\Omega _{m}+\Omega _{V}}{\left( 1+z_{BBN}\right) ^{6}\Omega _{m}^{2}}~.
\label{cond1}
\end{equation}%
The value of $\widetilde{\alpha }_{cr}$ is of the order of $10^{-24}$ for
reasonable cosmological parameters, thus we investigate only the negative $%
\widetilde{\alpha }$ range.

\begin{figure}[tbp]
\includegraphics[width=8.5cm, angle=0]{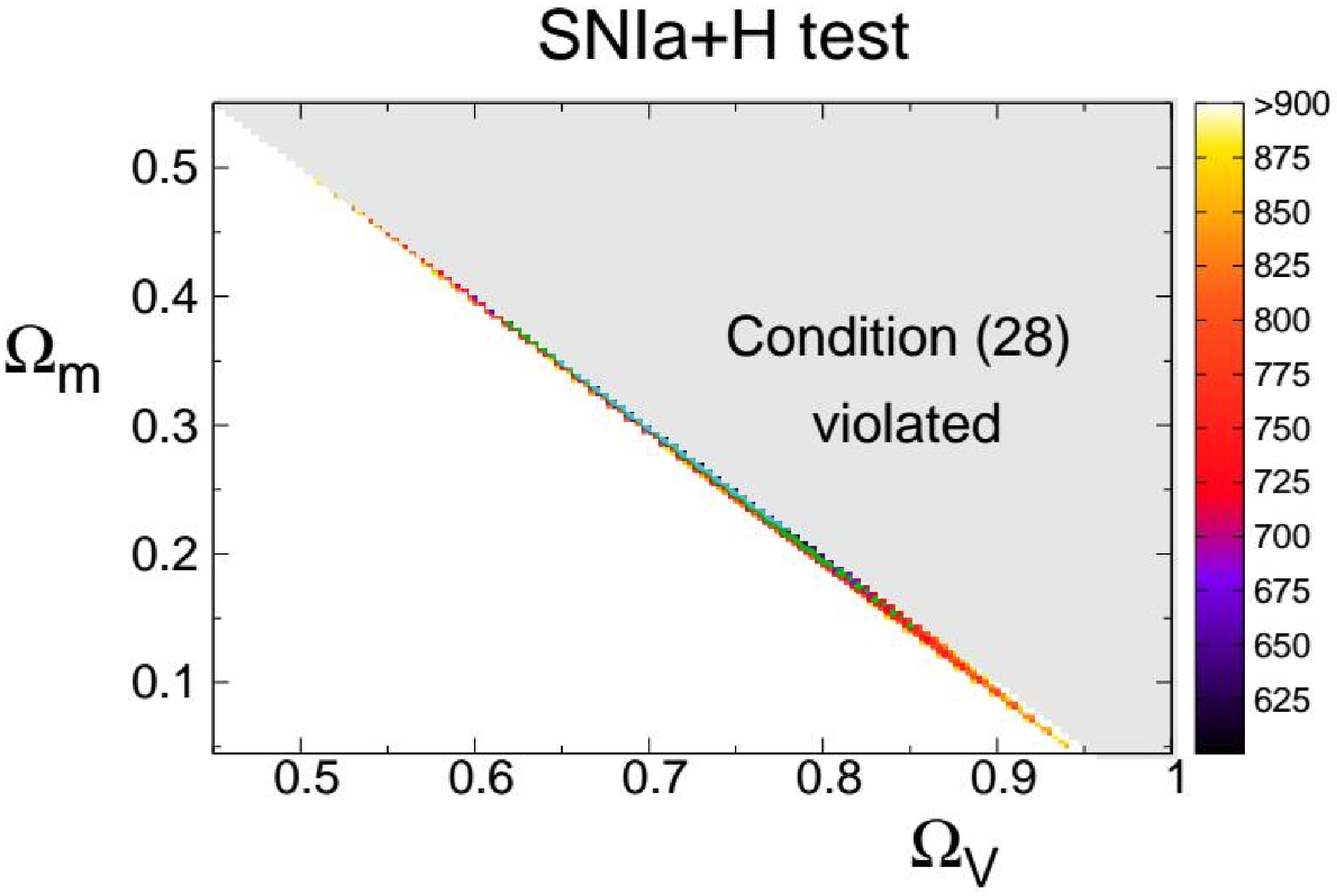}
\par
\vskip 0.5cm %
\includegraphics[width=8.5cm,angle=0]{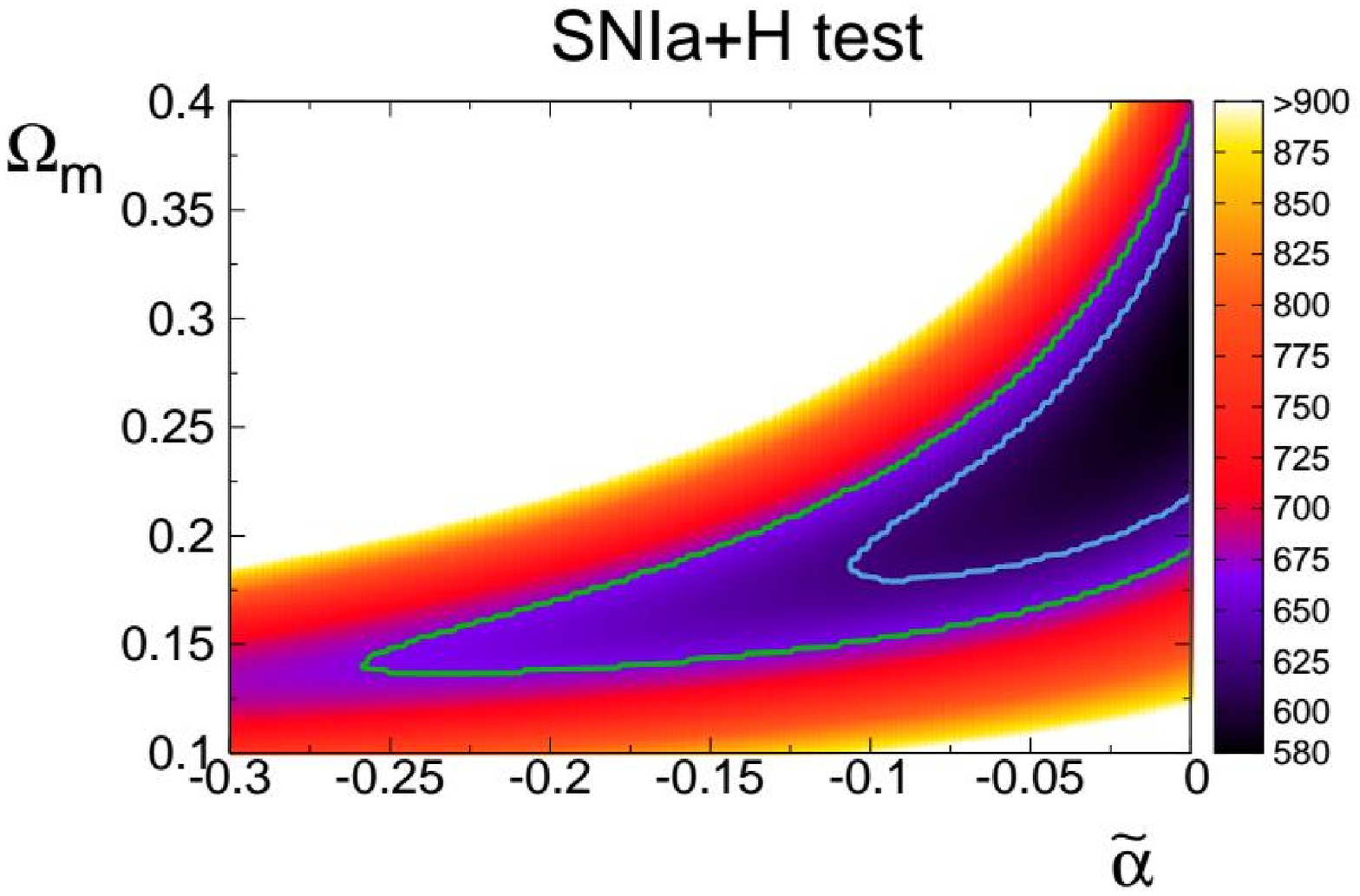}
\caption{(Color online) The test of the electric dark energy model with the
combined supernovae+Hubble parameter data set is shown. The color scale
represents the $\protect\chi ^{2}$ values on the parameter spaces ($\Omega
_{V}$,$\Omega _{m}$) (upper panel) and ($\widetilde{\protect\alpha }$,$%
\Omega _{m}$) (lower panel). Within the gray area the condition (\protect\ref%
{cond1}) is violated. The light-blue and the green lines bounds the 1$%
\protect\sigma $ and 2$\protect\sigma $ confidence regions, respectively.}
\label{TestESDESNIaH}
\end{figure}

In the test of electric dark energy model the Hubble constant is fixed by $%
H_{0}=67.74$ km/sec/Mpc in agreement with the measurements of Planck
spacecraft on the anisotropies of the cosmic microwave background radiation (CMB)
(see the last column of Table 4 in \cite{Planck}). The results of the tests
with the Union 2.1 SNIa data set are shown on Fig.~\ref{TestESDESNIa} in the
parameter space $\left( \Omega _{m},\Omega _{V}\right) $. The color scale
represents the $\chi ^{2}$ values and the light-blue and the green lines
bounds the 1$\sigma $ and 2$\sigma $ confidence regions, respectively. The
grey area shows the region where a bounce occurs between the redshift values
$0$ and $z_{BBN}$, i.e. where $\widetilde{\alpha }>\widetilde{\alpha }_{cr}$
. Good fit with SNIa data set is obtained along a narrow inclined stripe,
which nevertheless contains the $\Lambda $CDM limit ($\widetilde{\alpha }=0$
), with the widely accepted values of $\Omega _{m}\approx 0.3$ for the dark
plus baryonic matter sector and $\Omega _{V}\approx 0.7$ for dark energy.

The test with the combined SNIa and Hubble parameter data set further
narrows the 1$\sigma $ and 2$\sigma $ confidence regions, as shown on the
upper panel of Fig.~\ref{TestESDESNIaH}. On the lower panel the result of
the test with the combined data set is also represented in the parameter
space $\left( \widetilde{\alpha },\Omega _{m}\right) $.

{\color{black} The full fit in the 3-parameter space $\left( \widetilde{\alpha },\Omega _{m},H_{0}\right)$
 in the case of electric type model gave local minima, among which the LCDM limit was the best fit (the extra
parameter $\widetilde{\alpha }=0$). Thus the electric type models favour the $\Lambda $CDM models.}

\subsection{Cosmological test of a magnetic dark energy model}

\begin{figure}[tbp]
\includegraphics[width=8.5cm, angle=0]{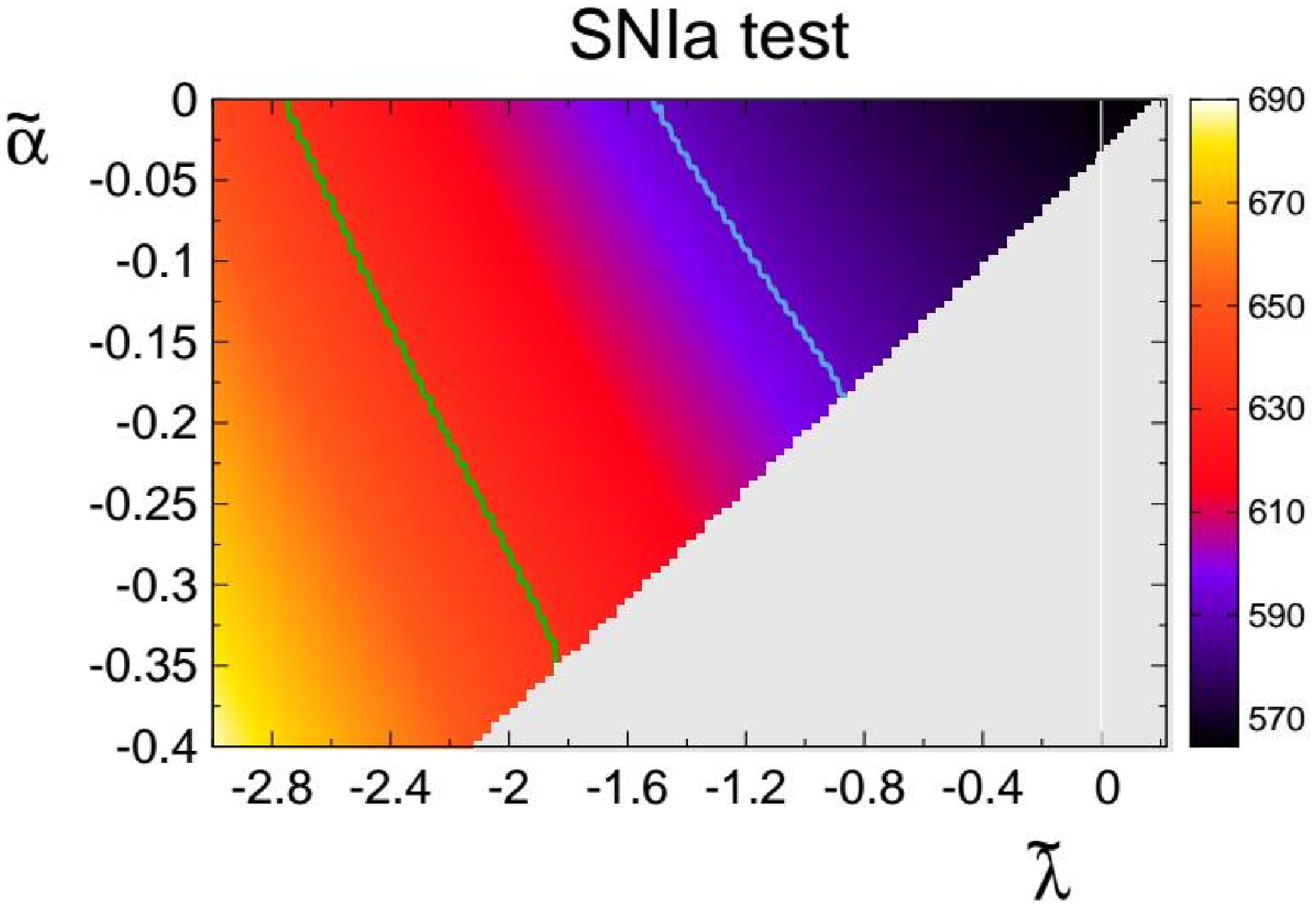}
\par
\vskip 0.5cm \includegraphics[width=8.5cm, angle=0]{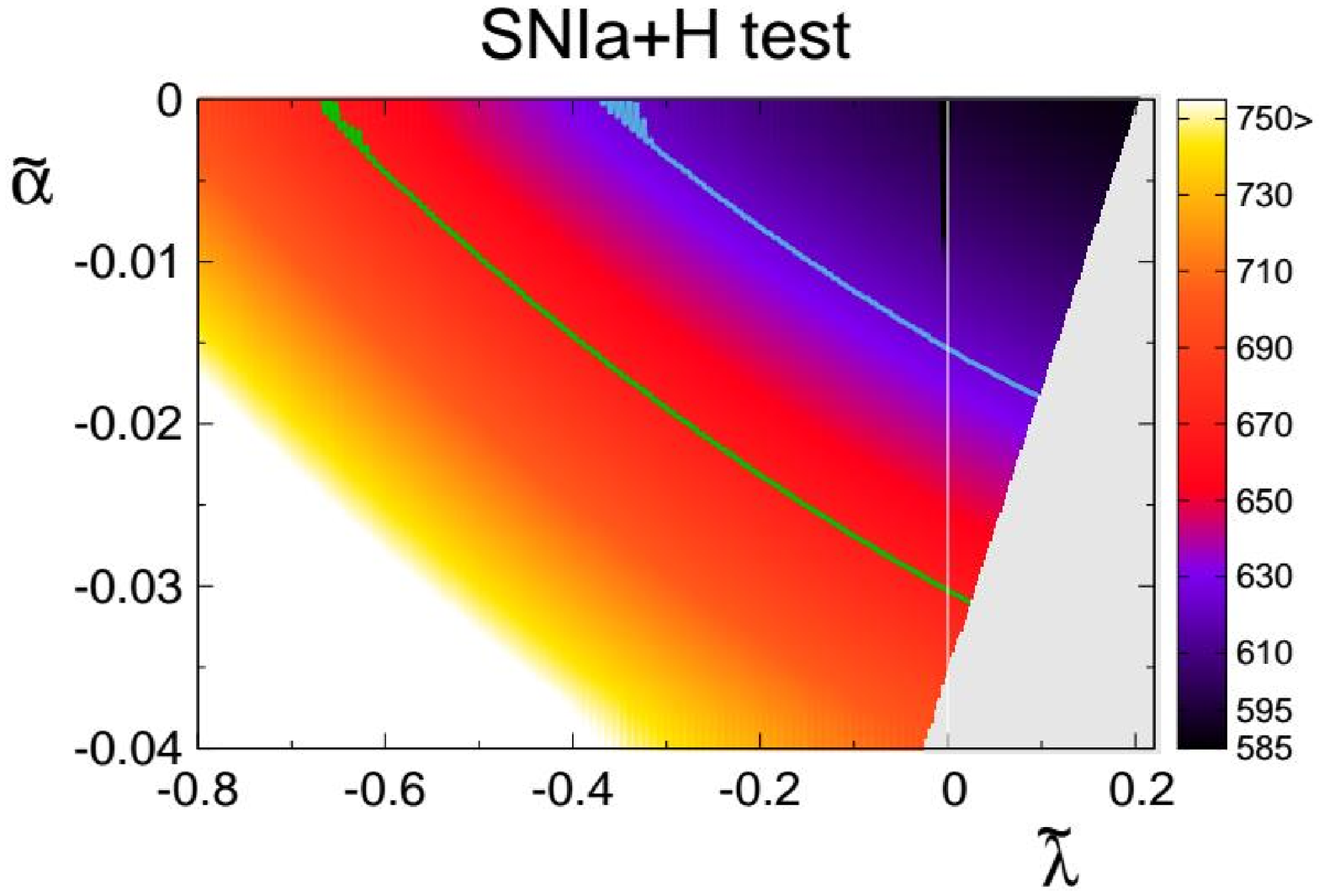}
\caption{(Color online) The test of the magnetic dark energy model with the
SNIa (upper panel) and the combined SNIa+Hubble parameter (lower panel) data
sets are shown. The color scale represents the $\chi ^{2}$ values on
the parameter spaces ($\widetilde{\lambda }$,$\widetilde{
\alpha }$). The light-blue and the green lines bounds the 1$\sigma $
and 2$\protect\sigma $ confidence regions, respectively. The fixed
parameters of the model are $\Omega _{m}=0.3088$, $\hat{A}_{0}=0.05$, $%
w_{DE}|_{z=0}=-0.99$, $H_{0}=67.74$ km/sec/Mpc and $d\hat{A}/dz|_{z=0}>0$.
The gray area gives the region where $\left( d\hat{A}/dz\right) ^{2}|_{z=0}$
would be negative for the chosen parameter values.}
\label{TestBSDESNIapH}
\end{figure}

In this model the dark energy vector potential is perpendicular to the four
velocity of the comoving matter. In accordance with the isotropy of the
assumed space-time geometry, all spatial components of the vector potential
are the same $A\left( z\right) $. {\color{black} From Eqs.~(\ref{52}) and (\ref{BHdot}), the Hubble parameter and its derivative are given by}
\begin{eqnarray}
\hat{H}^{2}\left( z\right) \left[ 1-\left( 1+z\right) ^{4}\left( \frac{d\hat{
A}}{dz}\right) ^{2}\right] &=&\Omega _{m}\left( 1+z\right) ^{3}+\Omega _{V}
\notag \\
&&\!\!\!\!\!\!\!\!\!\!\!\!\!\!\!\!\!\!\!\!\!\!\!\!\!\!\!\!\!\!\!\!\!\!\!\!\!%
\!\!\!\!\!\!\!\!\!\!\!-\widetilde{\alpha }\Omega _{m}^{2}\left( 1+z\right)
^{6}+\widetilde{\lambda }\left( 1+z\right) ^{2}\hat{A}^{2}~,  \label{HzMSDE}
\end{eqnarray}
{\color{black}
\begin{eqnarray}
\frac{\hat{H}\left( z\right) }{1+z}\frac{d\hat{H}\left( z\right) }{dz} &=&
\frac{3}{2}\Omega _{m}\left( 1+z\right) -3\widetilde{\alpha }\Omega
_{m}^{2}\left( 1+z\right) ^{4}+\widetilde{\lambda }\hat{A}^{2}  \notag \\
&&+2\left( 1+z\right) ^{2}\hat{H}^{2}\left( z\right) \left( \frac{d\hat{A}}{
dz}\right) ^{2}~,  \label{HderivBSDE}
\end{eqnarray}
}
where $\hat{A}\left( z\right) =A\left( z\right) /\sqrt{8\pi }a_{0}$ and%
\begin{equation}
\widetilde{\lambda }=\frac{4\pi \lambda }{H_{0}^{2}}~. 
\end{equation}%
The evolution of the dark energy vector potential is governed by Eq.~(\ref{54n}), and can be formulated as a function of the redshift:
\begin{equation}
\frac{d^{2}\hat{A}}{dz^{2}}+K\frac{d\hat{A}}{dz}+\widetilde{\lambda }\frac{
\hat{A}}{\left( 1+z\right) ^{2}\hat{H}^{2}}=0~,  \label{AEqBSDE}
\end{equation}
where
\begin{eqnarray}
K &=&\frac{3\Omega _{m}}{2}\frac{\left( 1+z\right) ^{2}}{\hat{H}^{2}}-3%
\widetilde{\alpha }\Omega _{m}^{2}\frac{\left( 1+z\right) ^{5}}{\hat{H}^{2}}
\notag \\
&&+2\left( 1+z\right) ^{3}\left( \frac{d\hat{A}}{dz}\right) ^{2}+\widetilde{%
\lambda }\left( 1+z\right) \frac{\hat{A}^{2}}{\hat{H}^{2}}~.
\end{eqnarray}

Since $\hat{H}_{0}^{2}=1$, Eq.~(\ref{HzMSDE}) gives a constraint among $%
\left( d\hat{A}/dz\right) ^{2}|_{z=0}$, $\hat{A}_{0}$, $\Omega _{m}$, $%
\Omega _{V}$, $\widetilde{\alpha }$ and $\widetilde{\lambda }$. Taking into account the Hubble constant, 
the evolution of the superconducting dark energy dominated Universe depends on six parameters.
{\color{black} Since Eq. (\ref{HzMSDE}) is a first integral of the system (\ref{HderivBSDE}) and (\ref{AEqBSDE}), it is used to check the accuracy of the numerical solution.}

\begin{figure}[tbp]
\includegraphics[width=8.5cm, angle=0]{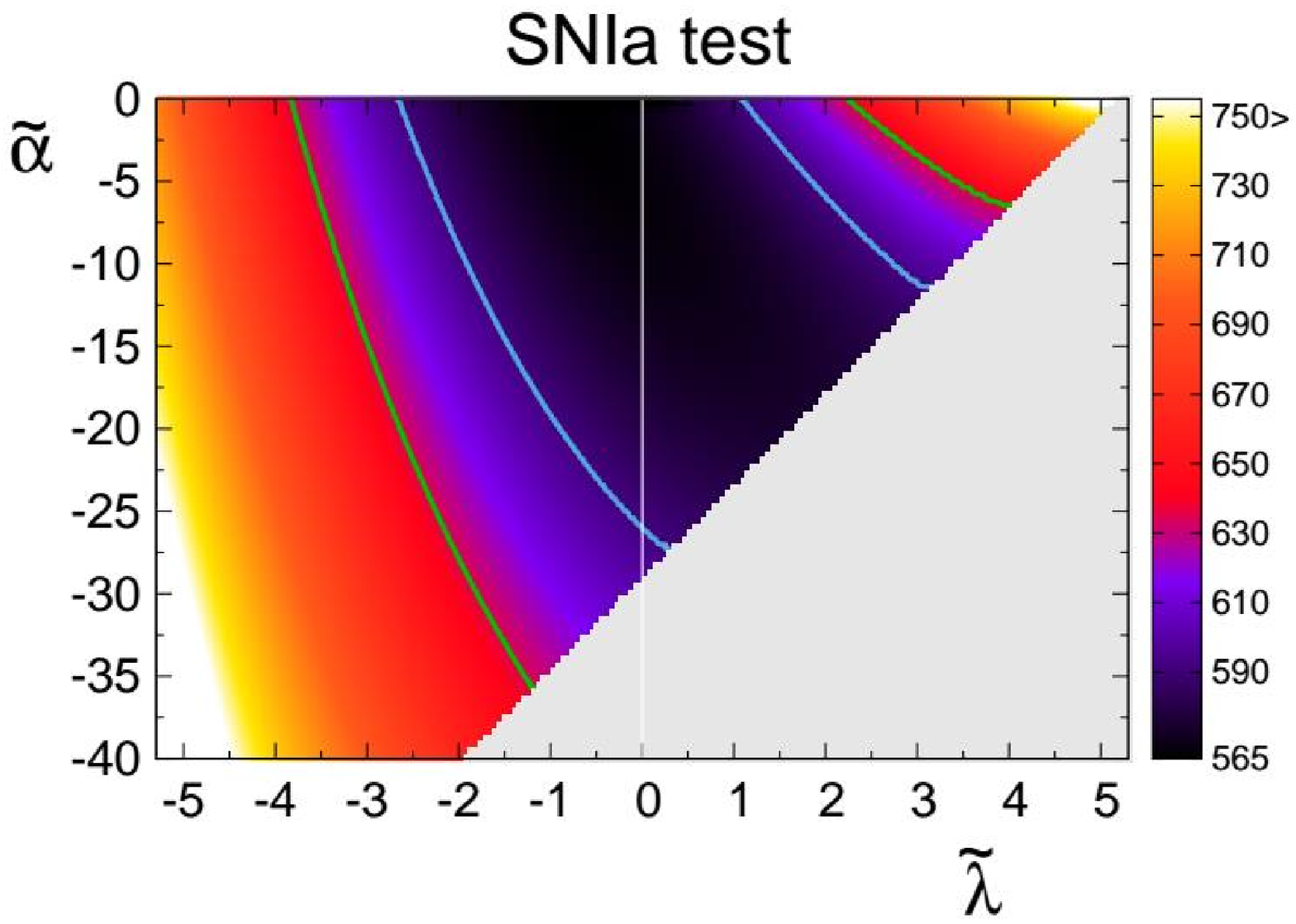}
\par
\vskip 0.5cm \includegraphics[width=8.5cm, angle=0]{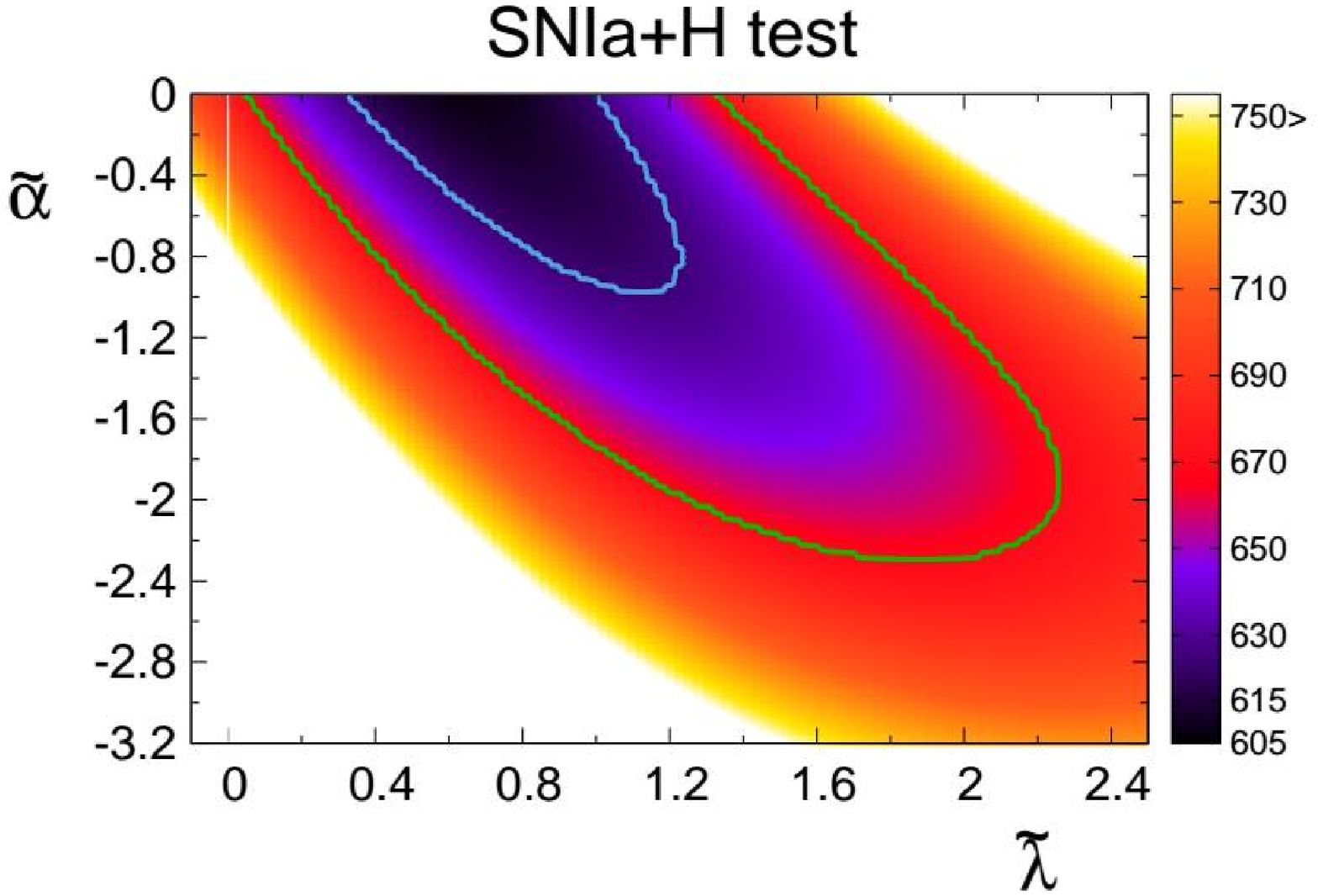}
\caption{(Color online) The same as on Fig. \protect\ref{TestBSDESNIapH} but
the fixed parameters are given by $\Omega _{m}=0.0486$, $\hat{A}_{0}=0.05$, $%
w_{DE}|_{z=0}=-0.855$, $H_{0}=70$ km/sec/Mpc and $d\hat{A}/dz|_{z=0}>0$.}
\label{TesttwoBSDESNIapH}
\end{figure}

The Hubble constant, and the cosmological parameter measuring the sum of baryonic and dark matter components, 
the matter density parameter are fixed as $H_{0}=67.74$ km/sec/Mpc and $\Omega _{m}=0.3088$, respectively. These
values emerge from the confrontation of the six-parameter $\Lambda $CDM
model with the observations of the Planck satellite (last column of Table 4
in \cite{Planck}). The parameter $w_{DE}$, giving the equation of state of
dark energy at present time, is chosen as%
\begin{eqnarray}
\left. w_{DE}\right\vert _{z=0} &=&\frac{\widetilde{\alpha }\Omega _{m}^{2}-
\frac{1}{3}\left( d\hat{A}/dz\right) ^{2}|_{z=0}+\frac{1}{3}\widetilde{
\lambda }\hat{A}_{0}^{2}+\Omega _{V}}{\widetilde{\alpha }\Omega
_{m}^{2}-\left( d\hat{A}/dz\right) ^{2}|_{z=0}-\widetilde{\lambda }\hat{A}
_{0}^{2}-\Omega _{V}}  \notag  \label{wDE} \\
&=&-0.99,
\end{eqnarray}
which is consistent with the Planck satellite's measurements \cite{Planck}.
We fix $\hat{A}_{0}$ as $0.05$, and choose $d\hat{A}/dz|_{z=0}>0$, such that
the function $\hat{A}\left( z\right) $ is decreasing at present. Then the
initial value $\left. \left( d\hat{A}/dz\right) ^{2}\right\vert _{z=0}$ can
be expressed as
\begin{eqnarray}
\hspace{-0.5cm}4\left( \frac{d\hat{A}}{dz}\right) ^{2}|_{z=0} &=&3\left(
1-w_{DE}\right) \widetilde{\alpha }\Omega _{m}^{2}  \notag \\
&&\!\!\!\!\!\!\!\!\!\!\!\!\!\!\!+3\left( 1+w_{DE}\right) G+\left(
1+3w_{DE}\right) \widetilde{\lambda }\hat{A}_{0}^{2}~,  \label{cond2}
\end{eqnarray}%
where%
\begin{equation}
G\equiv 1-\Omega _{m}+\widetilde{\alpha }\Omega _{m}^{2}-\widetilde{\lambda }
\hat{A}_{0}^{2}~.
\end{equation}
We note that Eq.~(\ref{cond2}) admits negative values for $\left. \left( d
\hat{A}/dz\right) ^{2}\right\vert _{z=0}$ for certain values of $\hat{A}_{0}$
, $\Omega _{m}$, $w_{DE}|_{z=0}$, $\widetilde{\alpha }$, and $\widetilde{
\lambda }$, which need to be excluded.

The results of the test of the magnetic dark energy model with the SNIa data
and with the combined SNIa+Hubble parameter data sets are shown in Fig.~\ref%
{TestBSDESNIapH} in the parameter space $\left( \widetilde{\lambda },%
\widetilde{\alpha }\right) $). We note that a bounce would occur between the
redshifts $0$ and $z_{BBN}$ for $\widetilde{\alpha }>0$ (except with the
infinitesimal positive values as in the electric dark energy model), and
therefore the Figure shows only the allowed parameter range $\widetilde{
\alpha }<0$.

We may also ask if the magnetic type superconducting dark energy fluid can
mimic the effects of both dark energy and dark matter, at least regarding
the data set on SNIa and Hubble parameter? In other words, is it possible to
have a unified dark energy and dark matter model? To answer this question we
fix the $\Omega _{m}$ parameter to the value $0.0486$ of the baryonic matter
alone \cite{Planck}. In order to reach good fit with the data, other
cosmological parameters could also change. We show the result of the test on
Fig.~\ref{TesttwoBSDESNIapH} for $w_{DE}|_{z=0}=-0.855$ and $H_{0}=70$
km/sec/Mpc as an example. As in the previous case, a bounce between the
redshifts $0$ and $z_{BBN}$ is avoided for $\widetilde{\alpha }<0$. It is
interesting to observe that the 1$\sigma $ and 2$\sigma $ confidence regions
are larger in this case.

The values of the free parameters of the model which are preferred by the
cosmological observations include a low value of the coupling between the
matter current and the fields ($\alpha \approx 0$), while the mass of the
dark energy vector field increases from $\lambda =0.01663$ $H_{0}^{2}$ to $%
\lambda =0.04584$ $H_{0}^{2}$ when it is also supposed to account for dark
matter.

\subsubsection{{\color{black} Best fit parameters}}

\begin{table}[h]
\caption{{\color{black} Some local minima of $\chi ^{2}$. The
coupling constant $\widetilde{\alpha }=0$ and $h_{100}$ is
defined by $H_{0}=100h_{100}$ km/s/Mpc.}}
\begin{center}
\begin{tabular}{l|lllll}
$\chi _{\min }^{2}$ & $\Omega _{m}$ & $w_{DE}$ & $\widetilde{\lambda }$ & $%
\hat{A}_{0}$ & $h_{100}$ \\ \hline
$578.76$ & $0.148$ & $-0.983$ & $4.568$ & $-0.030$ & $0.698$ \\
$578.99$ & $0.246$ & $-0.916$ & $0.175$ & $0.528$ & $0.697$ \\
$579.57$ & $0.268$ & $-0.997$ & $-0.009$ & $-0.373$ & $0.696$%
\end{tabular}%
\end{center}
\label{TableBSDE}
\end{table}

{\color{black} In this subsection we fix none of the six parameters of the
model. It is well-known that determining the global minimum of a function of six variables proves 
tremendously difficult due to the fact that any known algorithm for determining such a minimum could easily run 
into a local minimum. One of the most widely applied heuristic methods for a large parameter space is the Simulated Annealing Method. 
We have applied this procedure as implemented in Mathematica \cite{Mathematica} in order to find the minimum of the chi-squared function. 
The Simulated Annealing method lead to a multitude of local minima, suggesting 
severe degeneration in the parameter space, or in other words, elongated 1-sigma contours in the full parameter space.
The obtained smallest local minimum is $\chi ^{2}=578.76$.
A $\chi ^{2}$-difference test, as described in Ref. \cite{chisqdiff}, which showed that the fitting to the data sets is of closely similar 
quality for the investigated models as compared to the $\Lambda $CDM model. In other words, the $\Lambda $CDM model, arising as a special case in the family of models investigated, 
belongs to the 1$\sigma $ region of the parameter space.}

\begin{figure}[tbp]
\includegraphics[width=7cm, angle=270]{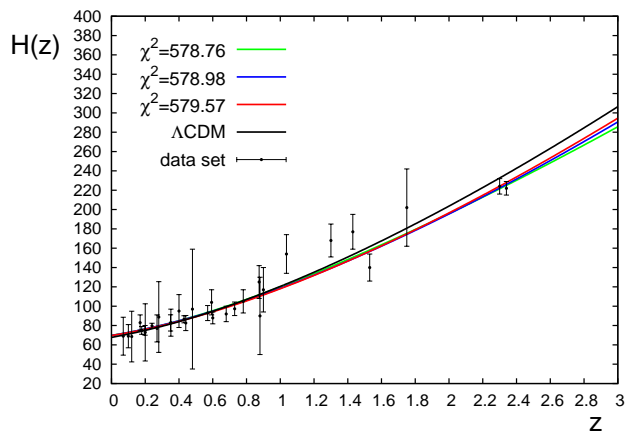}
\par
\vskip0.5cm \includegraphics[width=7cm, angle=270]{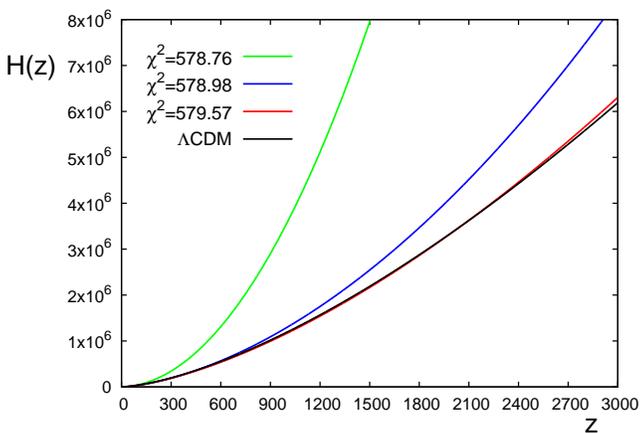}
\caption{(Color online) {\color{black} The evolutions of $H\left( z\right) $
in the magnetic type dark energy model (green, blue, red) and in the $
\Lambda $CDM model (black). Each curve represents local minima for $\chi ^{2}$ at points listed in Table \protect\ref{TableBSDE}. On the
upper panel the data set for Hubble parameter is also shown.}}
\label{HevoBSDE}
\end{figure}

\begin{figure}[tbp]
\includegraphics[width=7cm, angle=270]{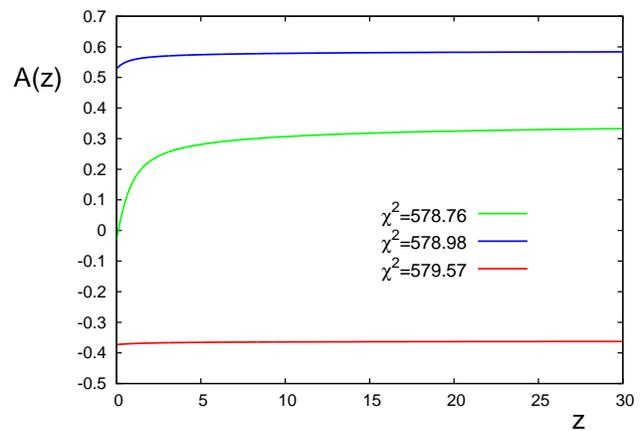}
\caption{(Color online) {\color{black} The evolutions of $\hat{A}\left(
z\right) $ related to the curves shown on Fig. \protect\ref{HevoBSDE}.}}
\label{AevoBSDE}
\end{figure}

{\color{black} Among the local minima we have found evolutions almost indistinguishable 
from $\Lambda $CDM, but also very different evolutions, with comparable fit. We represented the following local minima in Table \ref{TableBSDE} and on Figs. \ref{HevoBSDE} and \ref{AevoBSDE}: 
$i)$ the evolution belonging to the smallest local minimum found (green curve), 
$ii)$ the closest evolution to $\Lambda $CDM (red curve), and $iii)$ an evolution in between the two (blue curve). 
The evolutions of the Hubble parameter in the magnetic model (green, blue and red
curves) and for the $\Lambda $CDM model (black curve) are shown on Fig. \ref{HevoBSDE}. On the upper panel the
evolutions are represented for low redshifts and the data set for the Hubble
parameter are also given. The lower panel shows the evolutions for high
redshifts.
The evolutions of $\hat{A}\left( z\right) $ are represented on Fig. \ref{AevoBSDE}. In the distant past $\hat{A}\left( z\right) $
approximates a constant value, which is a typical behaviour in this model. }

\section{Concluding Remarks}

\label{concl}

In the present paper we have considered the cosmological tests of an
electromagnetic type dark energy model introduced in Ref. \cite{SupracondDE}
, in which the electromagnetic gauge invariance of the dark energy is
spontaneously broken. From a physical point of view the superconducting dark
energy model considered in our analysis is a two-field model, and extends
general relativity into a scalar-vector-tensor gravitational theory.
Alternatively, it can be seen as a unified scalar - vector field dark energy
model, in which the scalar field $\phi $ and the vector field $A_{\mu }$
appear in the gauge invariant, massive combination $\mathfrak{A}_{\mu
}=A_{\mu }-\nabla _{\mu }\phi $. {\color{black} From the point of view of cosmological applications, the present model may be used for realizations of early time inflation, with a possible explanation of the large scale CMB anomalies, or, as done in the present paper,  for describing  the late time acceleration of the Universe, and as a specific model of dark energy.

 An important theoretical test of gravitational models is the analysis of the
cosmological perturbations. The stability results also have important observational implications.  In vector-tensor models with action given by Eq.~(\ref{actv}), or in more general models with non-minimal coupling between vector fields and gravity, the analysis of the stability proved to be a
challenging task. In contrast to the scalar field gravitational models,  the linearized theory of vector-tensor models contains couplings between scalar, vector and tensor modes \cite{stab1}. It turns out that the perturbations decouple only in the ultraviolet limit \cite{stab1}, which, on the other hand, opens the possibility of the quantization of the models. Note that the
mixing of the different modes in vector - tensor models may lead to observable correlations of the scalar and tensor modes in the CMB.

In \cite{stab2} it was shown that a number of vector-tensor gravitational models contain instabilities in the form of ghosts, or unstable growth of the
linearized perturbations. These instabilities can be related to the longitudinal vector polarization modes presents in the theoretical models.
The presence of ghosts or tachyons in the early time/small wavelength regime of the vector-tensor cosmological models are an indication that the vacuum of the model is unstable. The instabilities found in \cite{stab2} growth up to horizon crossing, thus indicating
that the effective cosmological constant $\Lambda $ is comparable, or smaller
than the Hubble scale, implying  a different cosmological phenomenology. An analysis of perturbations in vector field models, with a special emphasis on inflation, was performed in \cite{stab3}, where the long wavelength limit of perturbations in small fields inflation and their linear stability was discussed. It turns out that the gravitational waves are unstable in large fields models. However, as a main result of this investigation, it was shown that one can use the small fields approximation in vector-tensor models as an effective theory for describing cosmological evolution.

The Hamiltonian stability and the hyperbolicity of vector field models involving both a general function of the Faraday tensor and its dual, as well as a Proca type  potential for the vector field, were considered in \cite{stab4}. The main result of this investigation is that the considered vector models do not satisfy the hyperbolicity conditions,  which require that the Cauchy problem is well posed. However, as shown in \cite{stab5}, in vector-tensor field theories with non-canonical kinetic terms, violations of hyperbolicity may not be present around the cosmological backgrounds.  Therefore vector models involving new cosmological dynamics and no causal pathologies can be found.

The problems of the stability of the linear perturbations and of the hiperbolicity is also of fundamental importance for the superconducting dark energy model. A full analysis can be done by considering the perturbations of the field equations Eqs.~(\ref{fn1}), (\ref{fn2}) and (\ref{DivF}), respectively, and analyzing their stability. However, we would like to point out that the superconducting dark energy model is explicitly constructed as a stable ground state of the U(1) gauge invariant scalar-vector field configuration. Hence we expect that this configuration is stable at least with respect to small linear perturbations, as is also the case for other similar vector-tensor models \cite{stab5}. 
}

We have considered the observational tests for two distinct classes of
models, whose physical and cosmological properties are determined by the
choice of the dark matter electromagnetic potential $A_{\mu }$. The first
model corresponds to an electric type dark energy potential, with $%
A_{0}\left( t\right) $ the only non-zero component, while in the second case
the space-like components of $A_{\mu }$ are different from zero, with $%
A_{\mu }=\left( 0,A(t),A(t),A(t)\right) $. For simplicity in both cases we
assumed a constant self-interaction potential for the electromagnetic type
dark energy with U(1) broken symmetry.

In the case of the electric model good fit with SNIa data set is obtained
along a narrow inclined stripe in the parameter space ($\Omega _{m},$ $%
\Omega _{V}$), which nevertheless contains the $\Lambda $CDM limit ($%
\widetilde{\alpha }=0$), with the widely accepted values of $\Omega
_{m}\approx 0.3$ for the dark plus baryonic matter sector and $\Omega
_{V}\approx 0.7$ for dark energy. The other points on this admissible region
represent essential deviations from the $\Lambda $CDM model (the black
region on the lower panel of Fig. 2), but they align along a narrow stripe
(upper panel of Fig. 2) with similar slope as the one given by the SNIa test
of the $\Lambda $CDM model, indicating that the dominant part of the
superconducting dark energy is still represented by a cosmological constant
type contribution, supplemented by a time-evolving component.

In the magnetic case the cosmological test select either i) parameter ranges
of the superconducting dark energy allowing for the standard baryonic plus
dark matter content $\Omega _{m}\approx 0.3$ or ii) a unified
superconducting dark matter and dark energy model, including additionally
only the baryonic sector $\Omega _{m}\approx 0.05$. A glance on Figs. 3 and
4, giving the admissible parameter spaces in the two cases shows that 

a) the
mass term $\lambda $ in the action has to increase in order to incorporate
dark matter in a unified model, and 

b) the cosmological data is best matched
for $\alpha \approx 0$, which decouples the matter current from both the
scalar and vector sectors of dark energy.

\section*{Acknowledgments}

Z. K. was supported by Hungarian Scientific Research Fund - OTKA Grant No. 100216. S.-D. L. gratefully
acknowledges financial support for this project from the Fundamental
Research Fund of China for the Central Universities.

\end{document}